\documentclass[aps,preprint,showpacs,superscriptaddress,groupedaddress]{revtex4-1}  
\usepackage{graphicx}  
\usepackage{dcolumn}   
\usepackage{bm}        
\usepackage{amssymb}   
\usepackage[usenames]{color}

\begin{document}

\title{L1$_0$ stacked binaries as candidates for hard magnets: FePt, MnAl and MnGa}

\author{Yu-ichiro Matsushita}
\affiliation{Max-Planck Institute of Microstructure Physics, Weinberg 2, D-06120 Halle, Germany}
\affiliation{Department of Applied Physics, The University of Tokyo, 
Tokyo 113-8656, Japan}
\author{Galia Madjarova}
\affiliation{Max-Planck Institute of Microstructure Physics, Weinberg 2, D-06120 Halle, Germany}
\affiliation{Department of Physical Chemistry, Faculty of Chemistry and Pharmacy, Sofia University, 1126 Sofia, Bulgaria}
\affiliation{Department of Physics, Indian Institute of Technology, Roorkee, 247667 Uttarkhand, India}
\author{C. Felser}
\affiliation{Max Planck Institute for Chemical Physics of Solids, N\"othnitzer Strasse 40, 01187 Dresden, Germany}
\author{J. K. Dewhurst}
\affiliation{Max-Planck Institute of Microstructure Physics, Weinberg 2, D-06120 Halle, Germany}
\author{S. Sharma}
\affiliation{Max-Planck Institute of Microstructure Physics, Weinberg 2, D-06120 Halle, Germany}
\author{E. K. U. Gross}
\affiliation{Max-Planck Institute of Microstructure Physics, Weinberg 2, D-06120 Halle, Germany}

\date{\today}

\begin{abstract}
A novel strategy of stacking binary magnets to enhance the magneto crystalline
anisotropy is explored. This strategy is used in the search for hard magnets by
studying FePt/MnGa and FePt/MnAl stacks. The choice of these binaries is
motivated by the fact that they already possess large magneto crystalline
anisotropy. Several possible alternative structures for these materials are explored
in order to reduce the amount of Pt owing to its high cost.
\end{abstract}

\maketitle

\section {Introduction}

The importance of magnets for modern society can hardly be overstated;
they enter into every walk of life from medical equipment to transport
(trains, planes, cars) to electronic appliances
(from household use to computers). All these devices use what is known as
hard magnets. The most common examples of such hard magnets are Nd$_2$Fe$_{14}$B,
SmCo$_5$ and Sm$_2$Co$_{17}$. These materials are the
strongest permanent magnets known to date.  Rare-earths are, however,
expensive and extracting them from mined ore is an highly polluting process.
It is therefore highly desirable to find a new generation of hard magnets that
contain less or no rare-earth atoms\cite{1,2,3,4}.

The first natural question to ask for designing new hard magnets is ``what
makes the existing permanent magnets hard from a microscopic point of view?''
For a magnet to be useful it needs to have two qualities (a) large saturation
magnetization density (Ms) and (b) a large magneto-crystalline anisotropy (MCA).
A Large MCA is needed to make a magnet stable w.r.t external
influences such as magnetic fields.
The large magnetization density in these rare-earth magnets is provided by the
ferromagnetic coupling between the magnetic atoms
(like Fe, Co or Sm), while the large MCA is due to both to the low symmetry of
the crystal structure and the large spin-orbit coupling provided by the
localized $f$-electrons. In order to design rare-earth free hard magnets one
needs primarily to rely on low symmetry of the crystal structure to achieve
large MCA since the spin-orbit coupling is considerably smaller in
$d$-electron materials.

There exist several strategies to design new hard magnets; in the
present work we explore a novel idea: step 1. identify two existing binary
compounds with desirable magnetic properties (with large Ms and
crystallizing in structures with high MCA); step 2. stack these
compounds together to form super-structures leading to further enhancement in
the MCA; and finally step 3. identifying the expensive component of this stack
and attempt to reduce it. This combination of the two binaries reminds one of
the quaternary Heuslers of the type XX'YZ where atoms X, X' and Y are $d$-elements,
the atom Z is a $p$-element with metallic
character\cite{5,6,7,8,9,10}.
Quaternary Heusler structures are mainly investigated as half-metallic
ferromagnets\cite{11,12,13,14,15,16} and are yet to
be investigated from the viewpoint of hard magnetic materials.

In the present work we explore this strategy by identifying FePt and MnAl
(or MnGa) as the binary compounds. These materials have L1$_0$ structure with
large MCA\cite{17,18}. The minimum energy and structures close-by for
 the stacks of these L1$_0$ magnets are then explored for ferromagnetically
coupled materials with large Ms. The most expensive component of these
compounds is Pt. We then look for possible  ways to reduce or replace Pt in the stacks. 

\section {Computational details}
Structural optimization for all materials was performed using the Vienna
ab-initio Simulation Package (VASP) with PAW pseudopotential
method\cite{19,20} using PBE exchange-correlation potential\cite{PBE}.
The energy cutoff of 350 eV and a Monkhorst-Pack 10x10x10 k-point grid was
used for all calculations. The possible magnetic phase space for a given
structure was explored using the highly accurate all-electron full-potential
linearized augmented-plane wave (FP-LAPW) method as implemented within the
Elk code\cite{21}. 

\section {Results and Discussion}
\begin{figure}
\includegraphics[width=0.4\linewidth]{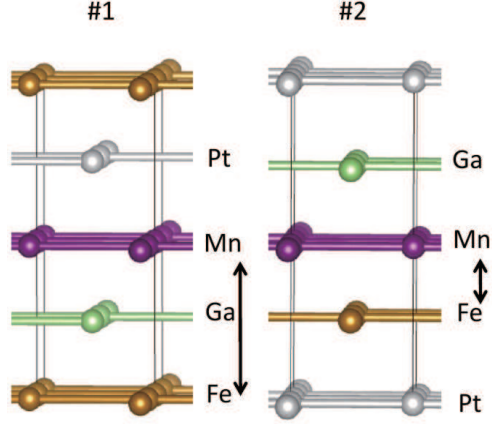}
\caption{Two possible 2xL10 structures of FePtMnGa. Different stacking
 sequences cause many possible polytopes, e.g. in the stacking sequence of the
 left side (FePtMnGa-stacking) magnetic layers of Mn and Fe are separated by
 non-magnetic layers in contrast to the right side structure (PtGaMnFe-stacking)
 where magnetic layers are next to each other.}
\label{2xL10}
\end{figure}

As a first step we look at possible structures for compounds FePtMnAl and FePtMnGa: 
(a) three quaternary Heusler structures\cite{22,23} based on the different
positions of the four atoms (see Table \ref{table1} for details) 
(b) two stacked L1$_0$ structures (which we call 2xL1$_0$ in the rest of the paper). 
These two structures are (see Fig.~\ref{2xL10}) built from four different layers
stacked in different order: in structure 1 each magnetic layer is separated by a
non-magnetic layer, while in structure 2 magnetic layers are 
nearest neighbors. These two 2xL1$_0$ structures are stoichiometrically identical
to the quaternary Heusler structures.

\begin{table}[htb]
  \begin{tabular}{|c|c|c|c|c|}
  \hline
F-43m & 4a & 4c &  4b & 4d\\
\hline
& (0,0,0) & (1/4,1/4,1/4) & (1/2,1/2,1/2) & (3/4,3/4,3/4)\\
\hline
type~I& Z & X' & Y & X\\
\hline
type~II& Z & Y & X' & X\\
\hline
type~III& Z & Y & X & X'\\
\hline
 \end{tabular}
  \label{table1}
  \caption{Three different types of atomic arrangement in the quaternary
  Heusler compound XX'YZ with the space group $F$-$43m$.}
\end{table} 

The calculated equilibrium lattice parameters, relative total energies and
atom-resolved saturation magnetic moment are presented 
in Table \ref{table2} and \ref{table3}. It is clear from these results that for
all the materials, the most stable structure is type I (cubic structure) with
ferromagnetic coupling between the magnetic atoms. This result is in a good
agreement with previously investigated quaternary Heusler
components\cite{6,16}. We also explored the possibility of tetragonal and
hexagonal distortion of the structure type I, and found that it is stable and
does not show any such distortion. For both materials the calculated magnetic
moment is around 4$\mu_B$ per formula unit which is in good agreement with the
Slater-Pauling rule\cite{24,25}: $M_t= Z_t-24$ where $M_t$ is total spin moment 
and $Z_t$ is total number of valence electrons which is 28 for these materials.

Structural optimization of Heusler structures of type II (this structure is
what is known as D022) and III show that
tetragonal distortion is favorable. Here we can see big differences in the
properties of the two investigated materials: for FePtMnGa structures II and III
are only $\sim$75 meV higher in energy than the cubic structure I and have
ferromagnetic ordering; while for FePtMnAl this energy difference is $\sim$300 meV.
These results are particularly interesting because tetragonal distortion is
highly desirable as it leads to an increase in the MCA.
Unfavorable results were obtained for layered structure 1 and 2 for both
materials: these structures are very high in energy ($\sim$400 meV) and the
magnetic coupling is ferrimagnetic. These structures are thus not good
candidates for hard magnets.

\begin{table}[htb]
\begin{tabular}{|c|c|c|c|c|c|}
\hline 
& $c/a$ & $\Delta E$ [meV] &  m$_{\rm Fe}$ & m$_{\rm Mn}$ & m$_{\rm t}$ [$\mu_B$]\\
I type & 1.0 & 0 & 1.03 & 3.03 & 4.15 \\ \hline
II type & 1.34 & 75 & 2.57 & 3.05 & 5.79 \\ \hline
III type & 1.23 & 72 & 2.58 & 3.26 & 5.94 \\ \hline
2xL1$_0$ str 1 & 2.64 & 317 & -2.57 & 3.12 & 0.58 \\ \hline
2xL1$_0$ str 2 & 2.62 & 320 & 2.69 & -2.03 & 0.85 \\ \hline
\end{tabular}
\label{table2}
\caption{Calculated relative energies per formula unit $\Delta E$ given in meV,
 optimized lattice constant ratio $c/a$ and magnetic moments in $\mu_B$ for
 quaternary FePtMnGa structures. For details of layered structures
 2xL1$_0$ type 1 and 2 see Fig. \ref{2xL10}}
\end{table}

\begin{table}[htb]
\begin{tabular}{|c|c|c|c|c|c|}
\hline
 & $c/a$ & $\Delta E$ [meV] &  m$_{\rm Fe}$ & m$_{\rm Mn}$ & m$_{\rm t}$ [$\mu_B$]\\
\hline
type~I& $1.0$ & $0$ & $0.93$ & $2.97$ & $4.03$\\ \hline
type~II& $1.14$ & $303$ & $2.62$ & $2.74$ & $5.55$\\ \hline
type~III& $1.43$ & $273$ & $-2.34$ & $2.92$ & $0.59$\\ \hline
2xL1$_0$ str 1& $2.51$ & $853$ & $-2.34$ & $2.77$ & $0.46$\\ \hline
2xL1$_0$ str 2& $2.51$ & $394$ & $2.61$ & $-1.76$ & $1.01$\\ \hline
\end{tabular}
\label{table3}
\caption{Calculated relative energies per formula unit $\Delta E$ given in meV,
 optimized lattice constant ratio $c/a$ and magnetic moments in $\mu_B$ for
 quaternary FePtMnAl structures. For details of layered structures
 2xL1$_0$ type 1 and 2 see Fig. \ref{2xL10}}
\end{table}

\begin{figure}
\includegraphics[width=0.7\linewidth]{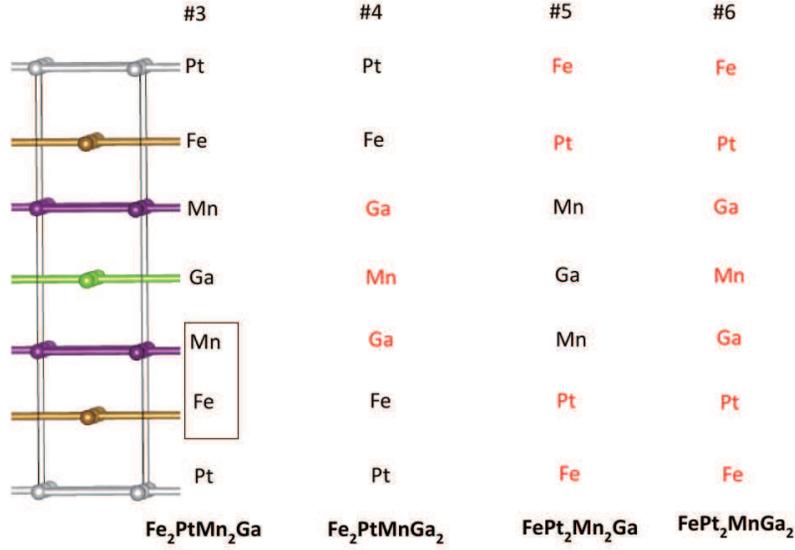}
\caption{Schematic of the various orders of stacked L1$_0$ structures
 types 3-6.}
\label{2xL10_1}
\end{figure}
We now focus upon the issue of reducing the amount of Pt. Two
possible routes to obtaining this are explored: (a) by a different type of
stacking and (b) by replacing Pt by less expensive metals. For the former we
examine four possible structure types the details of which are shown in Fig. \ref{2xL10_1}.

\begin{table}[htb]
  \begin{tabular}{|c|c|c|c|c|c|}
  \hline
 & m$_{\rm Fe}$ & m$_{\rm Mn}$ & m$_{\rm Mn}$ & m$_{\rm Fe}$ & m$_{\rm t}$ [$\mu_B$]\\
\hline
Fe$_2$PtMn$_2$Ga& $2.51$ & $-2.04$ & $-2.04$ & $2.51$ & $1.22$\\
\hline
Fe$_2$PtMnGa$_2$& $2.65$ & $-2.47$ & - & $2.47$ & $2.17$\\
\hline
FePt$_2$Mn$_2$Ga& $-2.96$ & $2.31$ & $2.31$ & - & $1.37$\\
\hline
FePt$_2$MnGa$_2$& $3.07$ & $2.46$ & - & - & $5.74$\\
\hline
   \end{tabular}
  \label{table3}
  \caption{Calculated magnetic moments in $\mu B$ for different layer structures of FePtMnGa.}
\end{table}

\begin{table}[htb]
  \begin{tabular}{|c|c|c|c|c|c|}
  \hline
 & m$_{\rm Fe}$ & m$_{\rm Mn}$ & m$_{\rm Mn}$ & m$_{\rm Fe}$ & m$_{\rm t}$ [$\mu_B$]\\
\hline
Fe$_2$PtMn$_2$Al& $-2.32$ & $1.69$ & $-1.69$ & $2.32$ & $0$\\
\hline
Fe$_2$PtMnAl$_2$& $2.41$ & $2.30$ & - & $-2.56$ & $2.07$\\
\hline
FePt$_2$Mn$_2$Al& $-2.96$ & $2.71$ & $2.71$ & - & $2.17$\\
\hline
FePt$_2$MnAl$_2$& $3.07$ & $2.28$ & - & - & $5.55$\\
\hline
 \end{tabular}
  \label{table3}
  \caption{Calculated magnetic moments in $\mu B$ for different layer structures of FePtMnAl.}
\end{table} 

Structure 3 is rich in Fe or Mn, and two different magnetic atoms are neighbors.
This configuration leads to ferrimagnetic ordering. Similar ferrimagnetic
ordering is also seen for structures type 4 and 5, in which magnetic atoms are
separated from each other by a layer of non-magnetic atoms. This renders all
these structure types inappropriate for the purpose of hard magnets.
The most interesting layered structure type turns out to be structure 6, where
the magnetic atoms are separated by two layers of non-magnetic atoms. In this
case, ferromagnetic ordering with a large magnetic moment is obtained. In order
to further reduce the amount of Pt we replaced it with Al, but kept the
structure rigid (i.e. no structural relaxation is performed). The magnetic
ordering and moment do not change substantially. However, if we perform
structural relaxation we find that the interlayer distance shrinks by 25\%.
This leads to ferrimagnetic ordering again as the lowest energy solution.
This is a very important finding: it is not the constituent atom but the
distance between the layers that plays the crucial role in determining the
magnetic order. To check the stability of the magnetic ordering of the original
material, i.e. FePtMnAl, we have further analyzed this structure. The effect
on the magnetic order and moment of the interlayer distances is examined.
These results are presented in Fig. 3 and indicate that $\pm$10\% of layer
distance does not change the stable magnetic ordering. If, however, one
reduces the distance further, the magnetic ordering changes to ferrimagnetic.
This suggests that Al is not a good choice but other atoms
with larger atomic radii should not result in such severe shrinking of the
interlayer distance.

\begin{figure}
\includegraphics[width=0.5\linewidth]{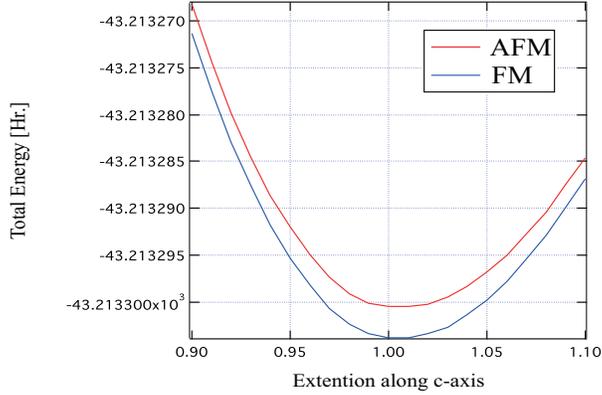}
\caption{Distortion of FePt2MnGa2 structure by changing the lattice constant.}
\label{2xL10_1}
\end{figure}

\begin{table}[htb]
  \begin{tabular}{|c|c|c|c|c|c|}
  \hline
 & m$_{\rm Fe}$ & m$_{\rm Mn}$ & m$_{\rm Mn}$ & m$_{\rm Fe}$ & m$_{\rm t}$ [$\mu_B$]\\
\hline
Fe$_2$AlMn$_2$Al& $0.79$ & $-0.46$ & $-0.46$ & $0.79$ & $0.6$\\
\hline
Fe$_2$AlMnAl$_2$& $-0.72$ & $-0.72$ & - & $2.26$ & $0.88$\\
\hline
FeAl$_2$Mn$_2$Al& $0$ & $2.25$ & $-2.25$ & - & $0$\\
\hline
FeAl$_2$MnAl$_2$& $-0.09$ & $1.11$ & - & - & $0.97$\\
\hline
   \end{tabular}
    \label{table4}
    \caption{Calculated magnetic moments in $\mu B$ for different layer structures of FePtMnAl.}
\end{table} 

\section{Summary}
We presented a novel approach for designing new hard magnets using stacks of
existing binary magnets. Such stacks could lead to high magneto
crystalline anisotropy owing to the lowering of crystal symmetry. We explored
this strategy using examples of FePt, MnAl and MnGa. We further tried to reduce
the expensive constituent of these stacks, i.e. Pt, by changing the
stoichiometry or by replacing Pt with a less expensive metal. In doing so we
identified two possible candidates namely FePt$_2$MnAl$_2$ and FeAl$_2$MnAl$_2$.
We suggest that these materials could be found experimentally and merit further
theoretical exploration. 

\section{Acknowledgments}
This work has been funded by the Joint Initiative for Research and Innovation
within the Fraunhofer and Max Planck cooperation program.

\end{document}